\documentclass[useAMS,usenatbib]{mn2e}
\usepackage{epsfig}
\usepackage{bethmacros}
\DeclareGraphicsExtensions{.eps, .jpg}

\def\spose#1{\hbox to 0pt{#1\hss}}
\def\lta{\mathrel{\spose{\lower 3pt\hbox{$\mathchar"218$}}
     \raise 2.0pt\hbox{$\mathchar"13C$}}}
\def\gta{\mathrel{\spose{\lower 3pt\hbox{$\mathchar"218$}}
     \raise 2.0pt\hbox{$\mathchar"13E$}}}

\title[MUGS]{Cosmological Galaxy Formation Simulations Using SPH}

\author[Stinson et al.]{
{G.\,S. Stinson$^{1,2}$\thanks{Email: gsstinson `at' uclan.ac.uk}, J. Bailin$^{1,3}$, 
H. Couchman$^1$, J. Wadsley$^{1}$, S. Shen$^{1}$, S. Nickerson$^{1}$, C. Brook $^2$,  
T. Quinn$^{4}$}
\vspace*{6pt}\\
$^{1}$Department of Physics and Astronomy, McMaster University, Hamilton, Ontario, L8S 4M1, Canada\\
$^{2}$Jeremiah Horrocks Institute, University of Central Lancashire, Preston PR1 2HE\\
$^{3}$Astronomy Department, University of Michigan,  830 Dennison Bldg., 500 Church St., Ann Arbor, MI 48109-1042\\
$^{4}$Astronomy Department, University of Washington, Box 351580, Seattle, WA, 98195-1580}

\begin{document}
\maketitle
\label{firstpage}

\begin{abstract} 
We present the McMaster Unbiased Galaxy Simulations (MUGS), the first 9 galaxies of an unbiased selection ranging in total mass from 5$\times10^{11}$ M$_\odot$ to 2$\times10^{12}$ M$_\odot$ simulated using n-body smoothed particle hydrodynamics (SPH) at high resolution.  The simulations include a treatment of low temperature metal cooling, UV background radiation, star formation, and physically motivated stellar feedback.  Mock images of the simulations show that the simulations lie within the observed range of relations such as that between color and magnitude and that between brightness and circular velocity (Tully-Fisher).  The greatest discrepancy between the simulated galaxies and observed galaxies is the high concentration of material at the center of the galaxies as represented by the centrally peaked rotation curves and the high bulge-to-total ratios of the simulations determined both kinematically and photometrically.  This central concentration represents the excess of low angular momentum material that long has plagued morphological studies of simulated galaxies and suggests that higher resolutions and a more accurate description of feedback will be required to simulate more realistic galaxies.  Even with the excess central mass concentrations, the simulations suggest the important role merger history and halo spin play in the formation of disks.
\end{abstract}


\section{Introduction}
\label{intro}
Forming a galaxy like our own Milky Way remains a challenge for the currently accepted $\Lambda$ Cold Dark Matter ($\Lambda$CDM) cosmogony.  The Milky Way is comprised of three distinct, stellar components: a flattened, rotating \emph{disk}; a compact, central and spheroidal \emph{bulge}; and a diffuse, spherical \emph{halo} of stars.  Any consistent cosmogony needs to explain the origin and evolution of each of these components.  $\Lambda$CDM posits that the energy budget of the Universe is currently dominated by vacuum energy ($\Lambda$), and the majority of the mass is invisible (dark) and only interacts with baryons via gravity.  Thus, in the early Universe, thermal baryonic pressure did not support the dark matter, and because it is non-relativistic (cold) the dark matter first collapsed into small structures.  Subsequently, the small structures merged hierarchically to form larger structures like the Milky Way.

The $\Lambda$CDM paradigm provides a simple explanation for the formation of the stellar halo:  stars formed early in small satellite galaxies, which got tidally stripped as their orbits brought them inside the tidal radius of the main galaxy.  While early observations indicated that stars in the Milky Way's halo might have condensed out of a monolithically collapsing gas cloud \citep{eggen62}, later observations found instead that formation through mergers like those proposed in the CDM paradigm are more likely \citep{searle78}.  Today, digital surveys of the sky reveal structures in the Milky Way's stellar halo such as streams and the remnant cores of dwarf galaxies \citep{Majewski1993,Belokurov2006}.  These are the exact signatures left by tidally disrupted satellites in simulations \citep{bullock05,abadi06}.  

Unlike halos, disks are not so neatly explained by $\Lambda$CDM.  Although conservation of angular momentum naturally creates rotating disks, the hierarchical buildup of structure impairs their formation, since disks form most efficiently in a quiet environment where gas cools and collapses smoothly.  In $\Lambda$CDM-inspired simulations of substructure mergers, satellites orbiting disks tidally heat stars turning thin disks into thick disks \citep{Toth1992,Quinn1993,Velazquez1999,Kazantzidis2008}.  Simulations of larger galactic mergers transform disks into centrally concentrated, spheroidal systems as the disks experience significant angular momentum loss \citep{Barnes1996,Cox2006}.  However, the observed distribution of galaxy morphologies can be reproduced with simple, analytic models.  In these models uniformly rotating spheres collapse into centrifugally supported disks \citep{Fall1980,Dalcanton97,mo98,vdB2001}.  The spheres start with angular momentum and mass profiles predicted using simulations of CDM structure formation.  The contradiction between mergers and disk formation may indicate that halo spin, not merger history, plays the dominant role in determining the morphology of galaxies.

The bulge is a spheroid of stars at the center of a galaxy.  The spheroidal shape indicates that they formed as the result of mergers.  However, other evidence reveals that some bulges may have a secular origin.  \citet{Kormendy1993} suggested that observations of rapidly rotating bulges indicates the existence of ``pseudo-bulges'', and recent simulations show that the central regions of isolated disks can buckle and cause stars to evolve through ``peanut'' shaped orbits into spheroidal distributions \citep{Debattista2004}.  

Whether disks or spheroids form affects many galactic properties like their kinematics, color, light distribution, star formation history and metallicity in addition to morphology.  Disk kinematics are dominated by ordered circular velocity, while in spheroidal components random velocity dominates over circular velocities.  Because star formation usually happens in galaxy disks where gas densities are sufficiently high, galaxies with more prominent disks have more recent star formation and display bluer colors, while spheroids tend to be more metal-rich and red.  The prominence of morphological components also has an impact on the radial distribution of galactic light profiles.  Galaxies with more prominent spherical components exhibit more centrally concentrated light profiles.  
\\
Models of galaxy formation require high resolution, hydrodynamic numerical simulations.  Analytic modeling can evolve a $\Lambda$CDM-motivated Gaussian density field into a spectrum of mass structures and populate those structures with stars such that they match the observered luminosity function of galaxies \citep{Cole2000, Benson2003, Somerville2008}.  However, because of the non-linear interactions of processes such as gas cooling, merging, tidal stripping, star formation, stellar feedback and active galactic nuclei, it is difficult for these models to predict galaxy morphologies, though \citet{Benson2009} and \citet{Dutton2009} represent recent attempts.  

As an alternative, numerical simulations allow us to study how halos, disks, and bulges were created and evolve.  Simulations that include gas are able to follow more physical processes than simulations that only track the gravitational interaction of dark matter.  Gas can be modeled using smoothed particle hydrodynamics (hereafter SPH), which partitions the gas in the Universe into particles and  with a Lagrangian approach follows the motions of those particles.  SPH is effective because it concentrates computational resources on the high density regions of a simulation, where galaxies form.  

Several studies of galaxies in a cosmological context have generated individual objects
that are similar to the observed local galaxies \citep{Governato2004, Robertson2004, Okamoto2005, Brook2006, Governato2007, Scannapieco2008, Scannapieco2009, Ceverino2009, SanchezBlazquez2009, Governato2009, Piontek2009, Martinez-Serrano2009}. These require high resolution and large computational resources in order for several important properties of the
simulations to converge, such as the galactic structure, 
motions and star formation
history.  As a consequence of the high computational cost, the
initial conditions have been carefully chosen to maximize the chance that the simulation will produce the desired type of object (usually a disk galaxy).

The previous cosmological simulations have shown that the collapse of gas is more complicated than smooth collapse into centrifugally supported disks.  They show that a significant fraction of gas flows into galaxies along cold filaments \citep{Keres05,Brooks2009}.  This raises the question: are centrifugal forces all that should support disks?  Observations suggest that disks are supported by an equipartition of energy between thermal, magnetic, and cosmic ray pressure support \citep{Cox2005}.

\subsection{Stellar Feedback}
One way to introduce pressure support into simulations is by harnessing the energy massive stars release in stellar winds and supernovae explosions.
Recent simulations continue to suffer from the overcooling that has long plagued morphological studies of simulated galaxies.  \citet{Navarro1991} described how angular momentum transferred from gas in the disk to the dark matter causes excess central mass concentrations, which leads to massive bulges that do not compare well to observations of disk galaxies whose bulges are fainter and less massive than their disks \citep{Allen2006}.  \citet{Navarro1991} instead proposed that stellar feedback could eliminate a significant amount of low angular momentum gas.  They implemented a method to kinematically excite gas particles around recently formed stars, but this showed little improvement in eliminating low angular momentum gas \citep{Navarro2000}.  

Stellar feedback plays a larger role in the development of satellite galaxies that merge with the parent galaxy.  The effects of feedback in satellites can contribute to the final morphology of the main galaxy.  If a satellite brings in more gas, it will contribute to make a larger disk; more stars will produce a larger spheroid.  In the main galaxies, stellar feedback may also determine the rate at which gas loses angular momentum and migrates through the disk into the dense, star forming center.

Due to insufficient resolution, there is currently no satisfactory treatment for stellar feedback.  Multiphase gas particles attempt to capture the phenomenology of the ISM inside individual particles \citep{SH03}, but it is difficult to determine the appropriate pressure for each particle to exert on the others and sometimes a stiff equation of state needs to be enforced \citep{Springel05}.  If the gas dynamics are separated into hot and cold gas \citep{Ritchie2001}, it is ambiguous when and how much gas should move from one phase to the other.  Disabling the gas cooling reproduces stellar feedback \citep{Gerrit97, TC2000, Stinson2006}, but resolution is often insufficient to turn off cooling in the proper amount of gas for the right length of time, and SPH does not allow single particles to create their own outflow.  Driven winds reproduce galactic outflows \citep{Navarro2000, Springel03, Oppenheimer2006, Okamoto2009}, but it is yet to be determined whether wind particles should be allowed to interact with surrounding gas and whether they provide sufficient pressure support to forming disks.  Many potential avenues need to be followed to see how each feedback recipe affects galaxy formation differently.  To date, no stellar feedback model has been effective at reducing the central mass concentration in high mass galaxies, though strong adiabatic feedback combined with altering the star formation threshold to a higher density in very high resolution simulations of low mass galaxies has had the most success \citep{Mashchenko2008,Governato2010}.

Computational resources can now support surveys that include a range of galaxies simulated at moderate (1 million dark matter particles inside $r_{vir}$) resolution \citep{Scannapieco2009,Okamoto2009,Piontek2009} with different stellar feedback recipes.  Each survey uses SPH, and each of the previous surveys have used \textsc{gadget}.  \citet{Scannapieco2009} separated hot and cold gas and used supernova feedback as a conduit from the cold to the hot phase.  \citet{Okamoto2009} used multiphase particles from \citet{Springel05} combined with driven winds of different strengths.  \citet{Scannapieco2009} found lower disk fractions than late type spirals in their simulations.  \citet{Piontek2009} tested several stellar feedback recipes and did not find striking success with any of them.

Observational samples of galaxies from galaxy redshift surveys, like
the 2dFGRS (Colless et al. 2001) and SDSS (York et al. 2000), now contain
millions of objects, allowing a much more complete view of not only the
properties of typical galaxies, but also a quantification of how galaxies are
distributed within the multivariate parameter space of galaxy properties.
In contrast, while many researchers are performing sophisticated galaxy
formation simulations, each can only produce a handful of galaxies.
Evaluating the success of these simulations requires a larger sample
of simulated galaxies. When only a small number of simulations exist,
it is easy to find a good observational match for any one simulation;
however, when a \emph{sample} of simulated galaxies exist that predict a mean and spread
for any galactic property, these predictions can be directly confronted
with the large observational samples.

In order to address these problems, we have begun the McMaster Unbiased
Galaxy Simulations (MUGS) project. The goal of MUGS is to generate a
large sample of sophisticated galaxy formation simulations that sample
potential sites of L* galaxy formation in an unbiased manner for
direct comparison to the large observational samples now available.
In this paper, we describe the methodology used for MUGS and
present an overview of the first 9 simulations, particularly focusing
on the relative formation of disks and bulges.

MUGS provides an extended look at galaxies simulated using similar physics to \citet{Governato2007} and \citet{Governato2009}.  Namely, supernovae are modeled with adiabatic ``blastwave'' feedback described in \citet{Stinson2006}.  In \S \ref{sec:sims}, we describe how we created the initial conditions for MUGS.  \S \ref{sec:code} details the algorithm that evolves the simulations. \S \ref{sec:results} examines the properties of the galaxies including their brightness, color, mass-to-light ratios, density profiles, bulge-to-total ratio, star formation history, and metallicity.

\section{Simulations}
\label{sec:sims}
The simulations that comprise the MUGS sample each use the volume renormalization techniques from many previous simulations.  The technique allows high resolution in a cosmological context at reasonable computational cost.  It focuses resolution in one specific region of a cosmological volume while simulating the rest of the volume at lower resolution.  The surrounding volume provides large scale density waves and impart tidal torques on the region of interest \citep{Quinn1992}.

We selected our galaxies from a cosmological cube 50 $h^{-1}$ Mpc on a side containing $256^3$ dark matter particles that was evolved to $z=0$.  The simulation is based on a 4096$^3$ realization of the CMBFAST \citep{Seljak1996} power spectrum initially subsampled to 256$^3$ to create a uniform resolution, dark matter-only volume.  It uses a WMAP3 $\Lambda$CDM cosmology with $H_0$ = 73 km s$^{-1}$ Mpc$^{-1}$, $\Omega_m$=0.24, $\Omega_\Lambda$=0.76, $\Omega_{bary}$=0.04, and $\sigma_8$=0.76 \citep{Spergel2007}.  

The uniform volume was evolved to $z$=0 at which point the friends-of-friends algorithm was used to find the virialized halos with a linking length = $\frac{1}{5}$ inter-particle separation \citep{Davis1985}.  Every group with a mass between $5\times10^{11}$ M$_\odot$ (705 particles in $256^3$) and $2\times10^{12}$ M$_\odot$ (2820 particles in $256^3$) was examined to ensure that it did not evolve closer than 2.7 Mpc  from any halo more massive than $5\times10^{11}$ M$_\odot$.  Massive structures contain hot gas that would significantly alter the evolution of a galaxy of interest.  Simulating structures larger than $2\times10^{12}$ M$_\odot$ at the resolution of these galaxies is currently computationally unfeasible.  Out of the 36,193 halos found with friends-of-friends, 761 were in the right mass range, and 276 of those were sufficiently isolated.  From that sample, 9 halos were selected using a random number generator for more detailed simulation without regard for spin parameter or merger history.  

\begin{figure}
\resizebox{9cm}{!}{\includegraphics{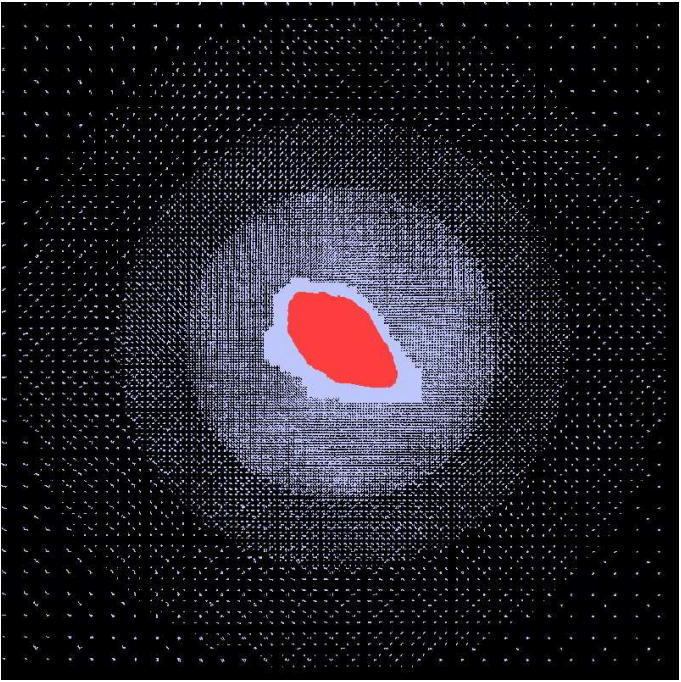}}
 \caption[Initial Conditions]{ Projection of initial particle configuration for g15784.  The red particles represent gas and the light blue represents dark matter.  Gas is only included in the high resolution region at the center.  The dots become less dense further away from the center because of the decreased resolution and higher particle masses at larger radii.}
\label{fig:ics} 
\end{figure}

Particles within 5$r_{vir}$ of each group's center at z=0 were traced to their positions in the initial conditions to specify the region of interest.  In an effort to efficiently use computational resources without loss of physical reality, the regions of interest have a non-spherical shape.  Minimizing the number of particles is critical since the n-body tree code must reconstruct the entire tree at each minor timestep, an operation that depends directly on the number of particles.  The central region was filled with a regular grid of particles to achieve an effective resolution of 2048$^3$ at the center.  Surrounding the non-spherical central region is a spherical region with a radius 1.2 times the maximum radius of the central region.  This immediately surrounding region is populated with particles for an effective resolution of $512^3$.  Outside this are three spherical regions equally spread in radius with effective resolutions of $256^3$, $128^3$, and $64^3$.  The outskirts of the 50 $h^{-1}$ Mpc cube are filled at an effective resolution of $32^3$.  Each of these regions contains progressively more massive particles corresponding to the reduced resolution.  Using this scheme, there are between 4 and 10 times more particles in the high resolution region than in the surrounding low resolution regions combined.  The regular grids of particles in each region were perturbed using the Zel'dovich approximation with subsampled force resolutions matching the particle resolutions.  This dark matter-only configuration was evolved to $z$=0.  The particles that ended up inside $3r_{vir}$ in the dark matter-only resimulation had a fraction of their mass converted into a separate, neighboring gas particle with a mass corresponding to the cosmic baryon fraction, $\Omega_b / \Omega_m$.  We then reapplied the Zel'dovich perturbations to this new regular grid.  Figure \ref{fig:ics} shows the initial configuration for a sample galaxy.  In the high resolution region, dark matter particles have a mass of $1.1\times10^{6}$ M$_\odot$ and gas particles have an initial mass of $2.2\times10^{5}$ M$_\odot$.  Stars form with a mass of $6.3\times10^{4}$ M$_\odot$.  Each particle uses a gravitational softening length of 312.5 pc.

\subsection{Galaxy Sample}
\label{sec:sample}
\begin{figure*}
\resizebox{18cm}{!}{\includegraphics{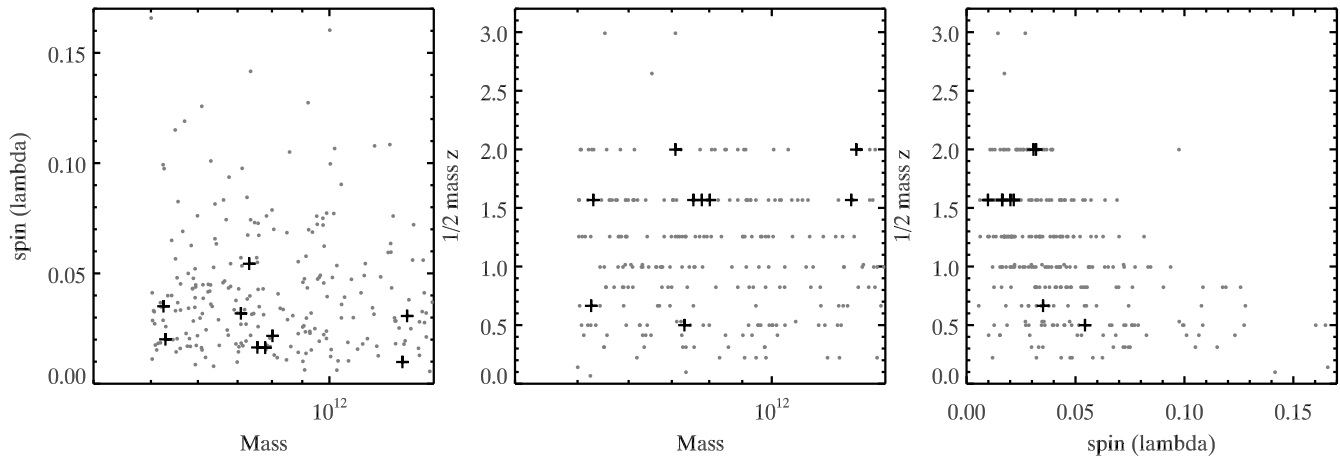}}
 \caption[Galaxy sample]{ Three projections of the galaxy sample in mass, spin, and half mass redshift space.  The light dots represent all the galaxies in our sample mass range from the uniform volume.  The plus signs indicate the galaxies presented here.}
\label{fig:sample} 
\end{figure*}

One of the main goals of this study is to develop a better understanding of what sorts of galaxy morphologies are created using a wide range of initial conditions.  The randomly selected MUGS sample spans a wide range of merger histories and halo angular momenta.  The merger history is characterized as the redshift at which the galaxy obtained half of its final mass.  The angular momentum of each halo is measured using $\lambda' = \frac{J}{\sqrt{5/3 G M^3 R}}$ \citep{Bullock2001}.  Figure \ref{fig:sample} shows how the $\lambda'$ and half mass redshifts of the halos in our 9 halo sample compare to the distribution of halos that passed our selection criteria.   The $\lambda'$ values shown in Figure \ref{fig:sample} are the values from the low resolution, uniform dark matter only simulation.  The values found when the galaxies are run at high resolution including baryonic physics vary modestly from these values.  Figure \ref{fig:sample} shows that in the initial sample of 9 galaxies does not include any from the high $\lambda'$ tail.  As the sample grows in the future, it will more fully reproduce the multivariate distribution of halo properties.  We note that only one halo with half mass redshift greater than 1.5 has a $\lambda'$ values greater than 0.05, but that some of the halos that accrete half of their mass later have $\lambda'$ values that surpass 0.1.  \citet{Ryden1988} point out that late accretion contributes more angular momentum to halos because of their larger turn around radius.  It is expected that galaxies with the quietest merger histories will form galaxies with the most prominent disks.

\section{Code}
\label{sec:code}
The simulations were evolved using the parallel SPH code \textsc{gasoline} \citep{gasoline}.  \textsc{gasoline} solves the equations of hydrodynamics, and includes radiative cooling.  Gravity is calculated for each particle using the \citet{Barnes1986} tree algorithm with tree elements that span at most $\theta$ = 0.7 of the size of the tree element's distance from
the particle.  \textsc{gasoline} is multistepping so
that each particle calculates its forces once every gravitational timestep
$\Delta{t_{\rm grav}}=\eta\sqrt{\frac{\epsilon_i}{a_i}}$, where $\eta$ =
0.175, $\epsilon_i$ is the gravitational softening length (312.5 pc), and $a_i$
is the acceleration.  For gas particles, the timestep must also be
less than $\Delta{t_{\rm gas}}=\eta_{\rm Courant}\frac{h_i}{c_i}$, where
$\eta_{\rm Courant}$ = 0.4, $h_i$ is the gas smoothing length and $c_i$ is the sound speed. 

The cooling is calculated from the contributions of both primordial gas
and metals as $\Lambda_{tot}(z, \rho, T, Z) = \Lambda_{HI, HeI, HeII}(z,\rho, T) +
\frac{Z}{Z_{\sun}}\Lambda_{metal,z_{\sun}}(z, \rho, T)$.  The primordial cooling follows the non-equilibrium evolution of internal energy along with three primordial gas species (HI, HeI, and HeII). H$_2$ cooling is not included. The scheme uses the collisional ionisation rates reported in \citet{Abel97}, the radiative recombination rates from \citet{Black81} and \citet{Verner96}, and bremsstrahlung and line cooling from \citet{Cen92}. The metal cooling grid is constructed using CLOUDY (version 07.02, last described by \citet{Ferland1998}), assuming ionisation equilibrium. A uniform ultraviolet ionizing background, adopted from Haardt \& Madau (in preperation; see \citet{HaardtMadau}), is used in order to calculate the metal cooling rates self-consistently.  The cooling lookup table is linearly interpolated in three dimensions (i.e., $\rho$, z, T) and scaled linearly with metallicity.  The energy integration independently uses a semi-implicit stiff integrator for each particle with the compressive heating and density (i.e. terms dependent on other particles) assumed to be constant over the timestep. 

The star formation and feedback recipes are the ``blastwave model" described in detail in \citet{Stinson2006} with additional improvements as described in \S~3.1.  They are summarized as follows.  Gas particles must be dense ($n_{\rm min}=0.1 cm^{-3}$) and cool ($T_{\rm max}$ = 15,000 K) to form stars.  A subset of the particles that pass these criteria are randomly selected to form stars based on the commonly used star formation equation, 
\begin{equation}
\frac{dM_{\star}}{dt} = c^{\star} \frac{M_{gas}}{t_{dyn}}
\end{equation}
where $M_{\star}$ is mass of stars created, $c^{\star}$ is a constant star formation efficiency factor, $M_{gas}$ is the mass of gas creating the star, $dt$ is how often star formation is calculated (1 Myr in all of the simulations described in this paper) and $t_{dyn}$ is the gas dynamical time.  The constant parameter, $c^{\star}$, is tuned to 0.05 so that the simulated Isolated Model Milky Way used in \citet{Stinson2006} matches the \citet{Kenn98} Schmidt Law, and then $c^\star$ is left fixed for all subsequent applications of the code.  This star formation and feedback treatment was one of the keys to the success of \citet{Governato2007} in producing realistic spiral galaxies in a cosmological simulation and the success of \citet{Brooks07} in matching the observed mass-metallicity relationship.

At the resolution of these simulations, each star particle represents a large group of stars (6.32 $\times 10^4$ M$_\odot$).  Thus, each particle has its stars partitioned into mass bins based on the initial mass function presented in \citet{Kroupa1993}.  These masses are correlated to stellar lifetimes as described in \citet{Raiteri96}.  Stars larger than 8 $M_\odot$ explode as supernovae during the timestep that overlaps their stellar lifetime after their birth time.  The explosion of these stars is treated using the analytic model for blastwaves presented in \citet{MO77} as described in detail in \citet{Stinson2006}.  While the blast radius is calculated using the full energy output of the supernova, less than half of that energy is transferred to the surrounding ISM, $E_{SN}=4\times10^{50}$ ergs.  The rest of the supernova energy is assumed to be radiated away.  Iron and Oxygen are produced in SNII according to the analytic fits used in \citet{Raiteri96}:

\begin{equation}
M_{Fe} = 2.802 \times 10^{-4} M_\star^{1.864}
\end{equation}

\begin{equation}
M_{O} = 4.586 \times 10^{-4} M_\star^{2.721}
\end{equation}

The iron, oxygen, and the supernova energy ejected from SNII are distributed to the same gas within the blast radius.  Each SNIa produces 0.63 $M_\odot$ Iron and 0.13 $M_\odot$ Oxygen \citep{Thielemann1986} and ejects it into the nearest gas particle for SNIa.

\subsection{Quantized Stellar Feedback}
One of the aspects of stellar feedback not given detailed consideration in \citet{Stinson2006} is the clustered nature of star formation.  In \citet{Stinson2006}, supernova feedback is chronologically distributed according to the stellar initial mass function and lifetimes.  The combination of Padua group stellar lifetimes with the \citet{Kroupa1993} IMF results in a constant, small energy release each feedback timestep (1 Myr, concurrent with star formation) for the 35 Myr until the lowest mass stars that explode as supernovae (8 M$_\odot$) explode.  Since the blastwave radius and cooling shutoff time are calculated from the energy released during a timestep, the \citet{Stinson2006} method results in small blastwaves.  \citet{McCray1987} describe how the accumulated energy of stellar winds and supernova feedback create large superbubbles around star clusters.  With a stochastic treatment of the energy release timing, we use the accumulated energy of all the stellar feedback that should result from a star particle to produce a larger, more realistic blastwave.  The larger blastwaves should provide more pressure support to make more extended disks.  Using a \citet{Kroupa1993} initial mass function, one supernova mass ($>$ 8 M$_\odot$) star is created for every 200 M$_\odot$ of stars that form.  Combining globular cluster \citep{Harris1996}, open cluster \citep{Piskunov2008}, and embedded cluster \citep{Lada2003} catalogues, we estimate that a typical star forms in a cluster of 4000 M$_\odot$.  Thus, we require a minimum energy release of 20 supernovae ($2\times10^{52}$ ergs) for the MUGS simulations.

We stochastically determine when a star particle releases feedback energy.  The probability that energy is released at each feedback timestep is 
\begin{equation}
p = \frac{(N_{SNII}~mod~N_{SNQ})}{N_{SNQ}}
\end{equation}
where $N_{SNII}$ is the number of supernovae calculated to explode during that star formation timestep and $N_{SNQ}$ is the ``supernova quantum'', the number of supernova required per explosion.  If the probability is greater than a random number selected between 0 and 1, $N_{SNQ}$ supernovae's worth of energy is released.  This causes SN energy to be released sporadically over the 35 Myr until the largest star remaining is $< 8$ M$_\odot$.

\section{Results}
\label{sec:results}
In this section, we examine the galaxies that form from our set of cosmological initial conditions.  First, we compare the properties of mock images of the simulated galaxies with observed galaxies.  Second, we examine which parameters have the greatest impact on how much disk or spheroid forms.  Third, we compare how much light is produced in our galaxies with observations of the Tully-Fisher relationship and weak lensing mass measurements. Finally, we present the star formation histories and metallicity trends for the simulated galaxies.

\subsection{Tabulated Results}
\label{sec:table}
In order to identify the main halo and its satellites, we used the Amiga Halo Finder (AHF)
\citep{Knollmann2009}. AHF overcomes the difficulties Friends-of-Friends has in separating
neighboring groups, by instead identifying density peaks using an adaptive mesh algorithm.  The adaptive mesh algorithm naturally leads to identification of substructure, which is critical for analyzing cosmological simulations.  Table \ref{tab:data} summarizes key properties for each of the galaxies found with AHF in the sample.  Galaxies are identified using the group number from the original friends-of-friends galaxy catalog.  The columns are described below.
\begin{table*}
\caption{Simulation data}
\begin{center}
\begin{tabular}{c|c|c|c|c|c|c|c|c|c|c|c|c|c}
Galaxy & Mass &$\lambda'$&$z_{lmm}$&$z_{1/2}$& $f_{b}$ &$M_{gas}$&$M_*$&$M_{disk}$&$M_{bulge}$&$N_{gas}$&$N_*$&$N_{DM}$& Gasless\\
 &($10^{11}$ M$_\odot$)& & & & &(all&masses&in&$10^{10}$ M$_\odot$)&$10^5$&$10^6$&$10^5$&  DM\\
\hline
g1536&7.0&0.025&2.9&1.8&0.159&5.1&6.0&1.8&3.9&2.4&1.4&5.3&1640\\
g5664&5.2&0.025&3.4&1.3&0.164&3.8&4.8&1.3&3.3&1.8&1.1&4.0&0\\
g7124&4.5&0.039&0&1.5&0.163&2.5&4.8&0.12&3.2&1.1&1.1&3.4&147\\
g15784&14&0.0345&3.42&1.4&0.150&10&11&3.3&5.5&4.8&2.6&11&2\\
g21647&7.7&0.048&0.18&0.53&0.168&5.8&7.1&1.1&3.3&2.7&1.6&5.8&6\\
g22437&8.8&0.013&1.56&1.2&0.170&7.6&7.3&0.56&5.6&3.5&1.6&6.6&1\\
g22795&8.7&0.011&0.032&1.3&0.144&6.0&6.4&0.26&5.4&2.7&1.5&6.6&76\\
g24334&7.7&0.041&0.085&1.2&0.162&7.1&11&1.1&5.4&3.3&2.4&8.2&60\\
g25271&13&0.014&2.9&1.3&0.147&8.9&11&2.2&7.2&4.0&2.4&10&3\\
\end{tabular}
\end{center}
\label{tab:data}
\end{table*}

The columns are defined as follows.  Mass is the total mass in $10^{11}$ M$_\odot$ located inside $r_{vir}$ at $z=0$ of the final simulation including baryonic physics.  $\lambda'$ is the spin parameter of that matter defined as $\lambda' = \frac{J}{\sqrt{5/3 G M^3 r_{vir}}}$.  $z_{lmm}$ is the redshift of the last major merger, which is defined as the redshift when the AMIGA halo finder no longer distinguished a satellite from the main halo where the satellite was greater than one-third the mass of the main halo at some point in its history.  $z_{1/2}$ is the redshift when the main halo was one-half its mass at $z=0$.  $M_{gas}$ is the mass of gas inside $r_{vir}$ at $z=0$.  $f_b$ is the baryon fraction of the final halo ($\frac{M_{star}+M_{gas}}{M_\mathrm{tot}}$).  $M_*$ is the mass of stars inside $r_{vir}$ at $z=0$.  $M_{disk}$ and $M_{bulge}$ are the mass of stars kinematically classified as part of the disk and bulge, respectively, as described in \S~\ref{sec:bd}.  All the gas and stellar masses are reported in $10^{10}$ M$_\odot$.  The numbers of gas, star, and dark matter particles inside $r_{vir}$ at $z=0$ are $N_{gas}$, $N_*$, and $N_{DM}$, respectively.

One of the problems inherent in running simulations where only a localized region is populated with high resolution particles is that it is possible for low resolution particles or particles from outside the gas region to pollute the region of interest and cause unphysical results.  No low resolution particles lie within $r_{vir}$ at $z=0$ of any of the MUGS simulations.  Some dark matter particles (``gasless DM'') that originated outside the gas region ended up inside the virial radius.  The presence of these particles without corresponding gas indicates that some of the gas in the simulation experienced slightly less pressure than in reality.  The initial conditions for g1536 were the first ones that were generated, before the final criteria were firmly in place, and only contained gas particle pairs for dark matter particles that ended up inside $r_{vir}$; as a consequence, it contains many more Bad DM particles than the other simulations.

\subsection{Simulated Images}
\label{sec:images}
\begin{figure*}
\resizebox{18cm}{!}{\includegraphics{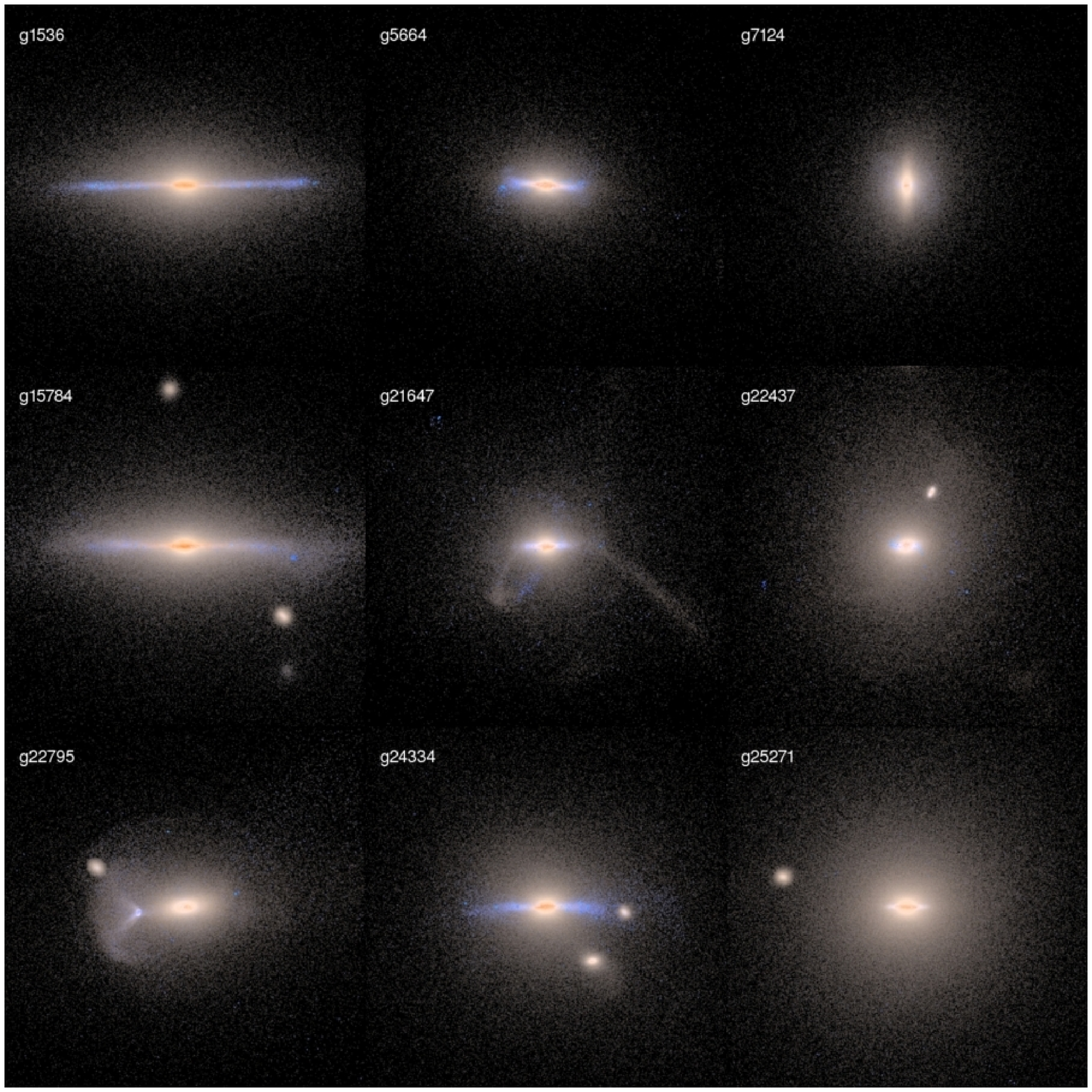}}
 \caption[Sunrise Images]{ Edge-on mock images of the simulated galaxies using the $g$, $r$, and $i$ filters found using the Monte Carlo radiative transfer program \textsc{sunrise}.  Each image is a 2D projection of a box 50 kpc on a side.}
\label{fig:sunrise} 
\end{figure*}

The most intuitive way to compare simulations with observations is through mock images of the simulations.  Such images can be created by assigning stellar population models like Starburst99 \citep{Leitherer1999} to each star particle to determine the color and luminosity each star particle should contribute to an image.  Additionally, dust can modulate the image with extinction and scattering.  \textsc{Sunrise} \citep{Jonsson06} is a Monte Carlo ray tracing program that produces simulated images by assuming dust exists in metal rich gas.  Figure \ref{fig:sunrise} shows images 50 kpc on a side that include scattering and absorption and are produced using \textsc{sunrise}.  The image brightness and contrast are scaled using $asinh$ as described in \citet{Lupton2004} since disks have an exponential surface brightness profiles meaning the images span a wide range in surface brightness. 

Each galaxy is aligned so that the total angular momentum of the gas inside 1 kpc is pointed upwards in Figure \ref{fig:sunrise}.  This presents the simulated galaxies edge-on to demonstrate the relative size of the disk and spherical component.  In several of the images, a thin disk of young, blue stars is surrounded by a halo of old, red stars.  In other images, little disk component is evident and the spherical component dominates.  We note that several of the galaxies (g5664, for example) are not perfectly aligned indicating that the disks are warped or that the stars are orbitting around a slightly different axis than the gas.  g7124 appears to be elongated vertically rather than horizontally because it is dominated by its spheroid which is strongly elongated perpendicular to its angular momentum of the inner gas, i.e. it is a spheroid rotating about the minor axis.

In contrast to the images presented in Figure \ref{fig:sunrise}, the magnitudes and surface brightnesses used throughout the rest of the paper are derived from the face-on projection of the galaxies generated by \textsc{sunrise}, for which the extinction effects are minimized.

\subsection{Color-Magnitude Relationship}
\label{sec:cmd}
\begin{figure}
\resizebox{9cm}{!}{\includegraphics{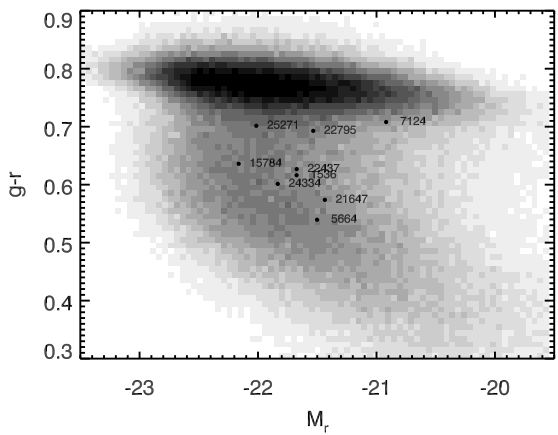}}
 \caption[Color Magnitude Diagram]{ Simulated galaxy $g-r$ color as a function of absolute $r$ magnitude in the Sloan Digital Sky Survey (SDSS) filter bandpasses overlaid on an inclination corrected color magnitude diagram of SDSS galaxies from $z < 0.2$ from \citet{Bailin2008}.  }
\label{fig:cmd} 
\end{figure}

One simple quantitative evaluation of the simulations is how the color
and brightness of galaxies compare to observations. Figure \ref{fig:cmd} shows a
$g-r$ versus absolute $r$ color-magnitude diagram (CMD) of $\sim 3 \times
10^5$ galaxies from the Sloan Digital Sky Survey (SDSS) as the shaded
two-dimensional histogram \citep{Bailin2008}. The
colors and magnitudes have been inclination-corrected to their expected
face-on values.
The two well-known features of the observed CMD are the relatively
narrow red sequence, which extends to very bright galaxies, and the more
broad blue cloud, which is abruptly truncated at $M_r \sim -23$. Relatively
few galaxies are observed to lie in the intermediate ``green valley''.
The simulated galaxies are overplotted as the labelled points.

The first conclusion that can be drawn from Figure \ref{fig:cmd} is that the simulated
galaxies lie in observationally-populated regions of the CMD: they are
representative of the colors and magnitudes of observed galaxies. Three
simulated galaxies, g7124, g22795, and g25271, lie on the red sequence while the
remainder are members of the blue cloud.  This is a slightly smaller proportion than the $48\%$ of SDSS galaxies that lie on the red sequence (within 0.08 of the mode of the red sequence, i.e. the \citet{Bailin2008} CMD parameter, $\mathrm{CMD^F} > -0.08$).
Given the small size of the simulated sample, this difference should not be
overinterpreted. However, a physical reason to expect different red
sequence fractions is that the two samples are found in different
environments: many observed galaxies are cluster members lying in massive
halos, while the simulated galaxies all lie at the centers of galaxy-mass
halos.

\begin{figure}
\resizebox{9cm}{!}{\includegraphics{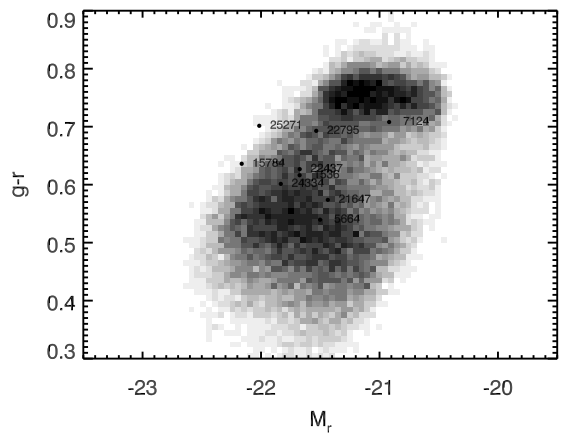}}
 \caption[Color Magnitude Diagram]{ As in Figure \ref{fig:cmd}, but with the SDSS sample restricted to isolated central galaxies with the same halo mass range as the simulated sample.}
\label{fig:isocmd} 
\end{figure}

This is demonstrated in Figure \ref{fig:isocmd}, where we have used the group catalogue of
\citet{Yang2007} to restrict the observational sample to
central galaxies of halos with masses $4 \times 10^{11}~M_{\sun} \le
M_{\mathrm{halo}} \le 2 \times 10^{12}~M_{\sun}$, and that do not have a
neighbouring group with halo mass $> 3 \times 10^{12}~M_{\sun}$ within
$500~\mathrm{km~s^{-1}}$ and a projected radius of $2$~Mpc. This is
essentially identical to the selection criteria for our simulated galaxies.
The relative fractions of blue versus red galaxies in Figure \ref{fig:isocmd} are
indistinguishable from those of our simulated sample: $30\%$ for the
observed sample and $3/9=33\%$ for the simulated sample.  It is also apparent from this
Figure that not only do the simulated galaxies lie within the observed
ranges of colour and magnitude for the global SDSS population, but they are also a particularly good match to observed galaxies that lie in halos of the
correct mass and environment (although perhaps somewhat too luminous, a point we will return to in Section 4.7).

While the MUGS galaxies are representative of some observed galaxies,
they are unusually ``green'': the red sequence galaxies lie on the blue
edge of the sequence, while the blue cloud galaxies lie in the red half
of the cloud. The simulated red sequence galaxies therefore have more
star formation activity than usually observed, while the blue cloud
galaxies
have either less recent or more ancient star formation than observed
galaxies. One explanation for the sustained star formation in the red sequence
galaxies is that it is a consequence of the environmental effects noted
above.  However, the simulated galaxies are also too blue in Figure \ref{fig:isocmd}, which
takes these environmental effects into account.  Their blue color could also highlight
some numerical failure of our simulations. MUGS does not include AGN
feedback, which might be able to drive significantly more star forming
gas out of the central regions of galaxies. The blue color may also
result from overcooling, where lack of resolution causes excess gas to
be driven to the galaxy center where it cools rapidly and forms stars
due to the high gas density.

The redness of the simulated blue cloud galaxies could also be a natural
result expected for moderate mass galaxies evolving in an isolated
environment, though Figure 5 again suggests that this is not the case.  Alternatively,
it could be the result of excess ancient star formation due to our
neglect of the increased feedback from metal-free Population III stars,
or overcooling of gas that resulted in too many stars formed in the
dense centers of galaxies at early times. Overcooling can in fact
simultaneously make the blue galaxies too red by building too large
of a bulge (as seen in \S \ref{sec:bd}), and make the red galaxies too
blue by continuing to provide cold gas to the centers of bulge-dominated
galaxies at late times.

Regardless of these details, the colors and magnitudes of the simulated
galaxies agree well with those of observed galaxies, giving us confidence
that analyzing them will provide a template of how galaxies form in the
real universe.

\subsection{Bulge vs. Disk}
\label{sec:bd}
\begin{figure}
\resizebox{9cm}{!}{\includegraphics{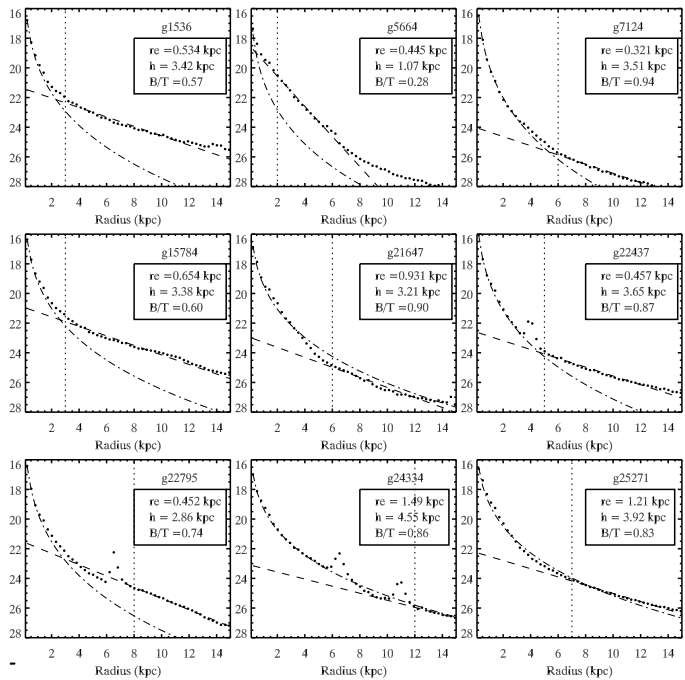}}
 \caption[Surface Brightness Profiles]{ Surface brightness profiles of face-on mock $i$-band images of the simulated galaxies, created using \textsc{sunrise}.  The profiles are fit with the sum of an $r^{1/4}$ law and an outer exponential disk.  The bulge-to-total ratios are calculated using equation \ref{eq:bt}.}
\label{fig:sbplots} 
\end{figure}

Another quantitative comparison between simulated galaxies and observations is how the light and matter are distributed.  Figure \ref{fig:sbplots} shows the face-on $i$ band surface brightness profiles of the simulated galaxies each fit with the sum of a de Vaucouleurs $r^{1/4}$ law (with effective radius, $r_e$) and exponential disk (with scale lengths, $h$) profiles.  We calculate a ratio of the light from the bulge to the total light of the galaxy using 
\begin{equation}
 B/T = \frac{I_e r_e^2}{I_e r_e^2 + I_d h^2}
\label{eq:bt}
\end{equation}
\citep{Binney1998}.  The calculated $B/T$ ratios shown in Figure \ref{fig:sbplots} are all higher than 0.5, indicating that bulges dominate the emission from all of our galaxies.  In contrast, observed $B/T$ ratios for galaxies with classical
bulge/disk profiles are often $<0.5$, (see, e.g., figure 8 of {Allen2006} base on the observed Millenium Galaxy Catalog).  Again, the high B/T ratios in the simulations indicate that too many stars form in the central region.

\begin{figure}
\resizebox{9cm}{!}{\includegraphics{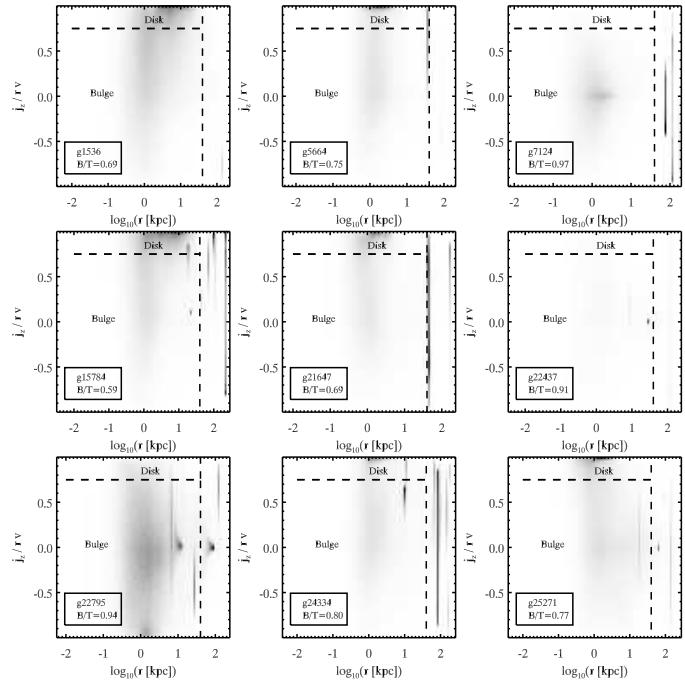}}
 \caption[Bulge-to-Disk Ratio]{ The ratio of the $z$-component of the specific angular momentum vector to the total specific angular momentum for every star particle as a function of its three dimensional radius from the center.  Stars are sorted into $100\times100$ bins 0.04 wide in log(r) and 0.02 high in $j_z r^{-1} v^{-1}_{\perp}$.  The horizontal dashed line at $\frac{j_z}{r v_{\perp}}$=0.75 indicates the separation between disk and spheroid.  The vertical dashed line at r=40 kpc indicates the radial extent inside which particles are classified as part of the spheroid.}
\label{fig:combjzr} 
\end{figure}

Galaxies can also be separated into bulge and disk components based on their kinematics. For this analysis, disks are aligned such that the angular momentum of the gas within 30 kpc of the center of the halo points along the z-axis.  Figure \ref{fig:combjzr} shows the ratio of the $z$-component of the specific angular momentum vector to the total specific angular momentum for every star particle, as a function of its three dimensional radius from the center.  Thus, Figure \ref{fig:combjzr} represents a comparison between the angular momentum of the gas and the angular momentum of the stars.  Each star within 40 kpc of the galactic center is classified as belonging to the disk if greater than 0.75 of its total angular momentum is aligned with the gas, or to the bulge otherwise.  The ratio of the masses of these components is given as the bulge to disk ratio.  Note, though this is not the same bulge-to-total ratio derived photometrically as in Figure \ref{fig:sbplots}, the results are similar as the mass of stars in the bulge is always greater than the mass in the disk.  Since the 40 kpc radius is much larger than what is generally considered part of the bulge, hereafter we refer to the kinematically defined component as the spheroid.

\subsubsection{What Creates the Disks?}
\begin{figure*}
\resizebox{18cm}{!}{\includegraphics{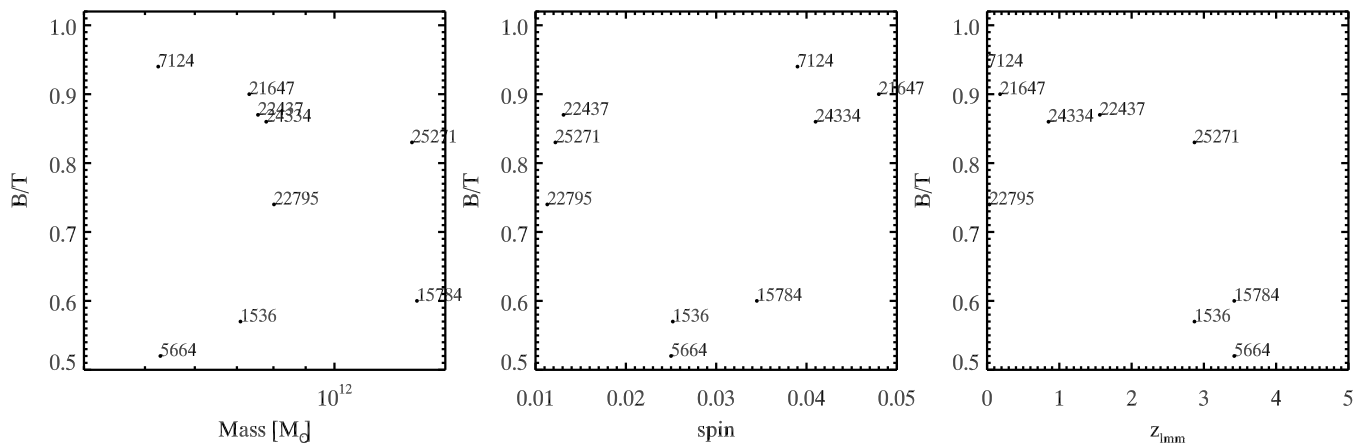}}
 \caption[Bulge-to-Total Comparison]{ The photometric bulge-to-total ratios shown in Figure \ref{fig:sbplots} as a function of galaxy mass, spin, and last major merger redshift.  Galaxies at the bottom of these plots exhibit the strongest disks although they
still have $B/T > 0.5$.}
\label{fig:bdcomp} 
\end{figure*}

The bulge/disk decompositions show that MUGS simulations remain crude representations of the formation of disk galaxies; the bulge fractions are much larger than what are observed.  Still, young stellar disks are apparent in at least g1536 and g15784 in Figure \ref{fig:sunrise}, so we briefly examine which halo properties correlate with disk formation in the simulations.  Figure \ref{fig:bdcomp} shows the photometric bulge-to-total ratio for the galaxies as a function of mass, final spin parameter ($\lambda'$), and the redshift of the last major merger.  The use of unbiased simulations allows us to draw conclusions
from these plots because no selection criteria was based on those
variables. One exception to our use of the photometric B/T ratio is g5664, which is
the only galaxy with significantly different photometric and kinematic B/T
ratios. The photometric decomposition for g5664 is unusual, and seems to
display two distinct exponential components. The total bulge+disk fit is
therefore poor and the decomposition based on that fit is suspect. Thus,
for g5664, we have plotted the mean of the kinematic and photometric B/T
ratios in Figure  \ref{fig:bdcomp}.  

A correlation is apparent between $z_{lmm}$ and B/T.  Galaxies with the most recent last major mergers have the largest bulges and typically the reddest colors in Figure \ref{fig:cmd}.  One galaxy that had its last major merger long ago, but still has a large bulge is g25271.  We note that g25271 has one of the lowest spin parameter values.  If we look in detail at the history of g25271, we find that it was involved in enough retrograde minor mergers that it never had a chance to grow a significant disk.

We note that the galaxies with the most significant bulge separate into two groups, one with higher spin and one with lower spin.  The higher spin galaxies all experienced recent last major mergers \citep[also see][]{D'Onghia2007}.  This indicates that it is only possible for galactic halos to gain angular momentum sufficient to increase its spin parameter above 0.04 if it has a late major merger.  We further note that that these galaxies with a high spin parameter display the largest bulge fractions.  Observed giant low surface brightness galaxies also exhibit large bulges \citep{Lelli2010}.  This might confirm the predictions of analytic models of galaxy formation \citep[e.g.][]{Dalcanton97} that halos with the highest spin parameters should host low surface brightness galaxies.  \citet{Governato2009} showed how disks can form after late major mergers, but refrained from making any direct comparisons to LSBs. 

The lower spin group experienced a wide range of merger histories.  Based on their spin parameters, it is apparent that they must have undergone a significant number of retrograde mergers.  

\subsubsection{What Creates the Spheroids?}
\begin{figure}
\resizebox{9cm}{!}{\includegraphics{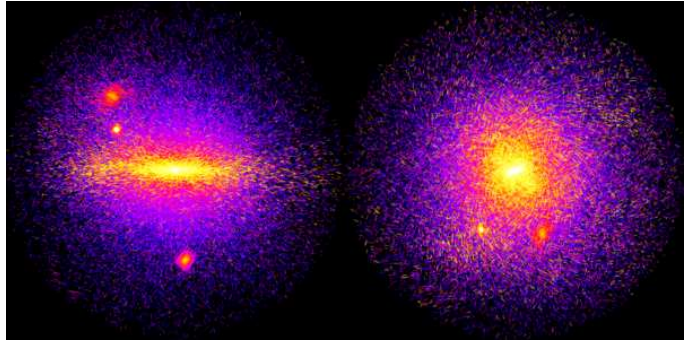}}
 \caption[Bulge Star velocities]{ Velocities of stars kinematically classified as part of the spheroid in g15784 at z=0 are shown as the thin lines.  The stars are colored by their age with more recently formed stars yellow and less recently formed stars blue.  The image is 80 kpc across.}
\label{fig:orbits} 
\end{figure}

Every galaxy in our sample displays a dominant spheroidal stellar component in both photometric and kinematic decompositions.  Since spheroids easily form as the result of merging, we examine the formation of the spheroid in the simulation that has the quietest merger history and largest disk (g15784) here to discover whether there are any significant secondary effects.  We leave study of spheroid formation in the other galaxies for future work.  Figure \ref{fig:orbits} shows the orbits of stars classified as spheroid stars in g15784 using the kinematic decomposition.  The stars are colored by age so that recently formed stars are yellow and stars that formed long ago are blue.  The edge-on view shows that many of the recently formed ``spheroid'' stars are actually orbitting in the disk plane.  The face-on view shows that many new stars are being formed in the central region of the galaxy with radial velocities that lead to their classification as ``spheroid'' stars.  While Figure \ref{fig:orbits} is colored to emphasize the young stars orbitting in a disk, the overall shape of the stars classified as part of the spheroid is spherical.  

\begin{figure}
\resizebox{9cm}{!}{\includegraphics{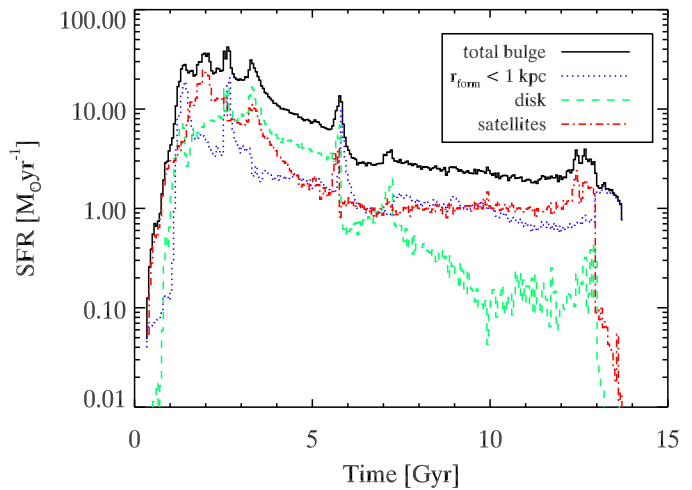}}
 \caption[Spheroid SFH]{ Formation history of stars kinematically classified as part of the spheroid in g15784 at $z=0$.  The \emph{solid line} shows the total star formation history for all the stars that comprise the spheroid at $z=0$.  The \emph{blue dotted line} shows the portion of that total that formed in the central 1 kpc of the main halo.  The \emph{green dashed line} shows the portion that formed in the disk and was heated to the spheroid.  The \emph{red dot-dashed line} shows the portion of the inner 40 kpc spheroid that formed in satellites. }
\label{fig:bulgesfh} 
\end{figure}

Figure \ref{fig:bulgesfh} shows the formation history of the stars in the spheroid at $z$=0.  The spheroid stars are separated into three categories based on where they formed, 1) in satellites; 2) within 1 kpc of the center of the main galaxy; 3) or in the disk of the main galaxy.  We emphasize that stars in the disk category are no longer part of the disk, but are spheroid stars that formed in the disk.  The decomposition was based on the distance stars formed from the center of the main halo.  The stars were sorted into bins one million years in length based on their formation time.  The center of mass of the stars that formed during that time was calculated by combining the stars formation locations in each time bin with the center of the main halo found using the AMIGA halo finder \citep{Knollmann2009} at the nearest output (generally full outputs were only written every 100 Myr).  The large number of stars that formed within the central kpc of the main halo made finding the star formation center straightforward for the MUGS galaxies. We calculated the three dimensional distance between the center and the
stars that formed during each time interval, and placed them into radial
bins 1 kpc wide. Because the stellar density drops strongly at the edge of
the disk, the existence of an empty radial bin serves as a delineation
between the main galaxy and satellites. We therefore classify all stars
within this radius (but outside of 1 kpc) as ``disk-formed'' stars, and all
stars outside of this radius as ``satellite-formed'' stars.  The total bulge stellar mass is 1.1$\times10^{11}$ M$_\odot$, the mass formed inside 1 kpc is 2.9$\times10^{10}$ M$_\odot$, the mass formed in the disk is 3.3$\times10^{10}$ M$_\odot$, and the mass formed in satellites is 4.1$\times10^{10}$ M$_\odot$.  

The final merger that disturbs stars from the disk into the spheroid happens about 13 Gyr into the simulation.  After that time, nearly all the stars that constitute the spheroid form in the central 1 kpc.  These are the stars that appear yellow in Figure \ref{fig:orbits}.  Stars forming in the unresolved central kpc is not limited to the final Gyr, they are a constant source of spheroid stars throughout the evolution of g15784.  For a significant fraction of the history of g15784, the mass of stars formed in the central kpc of the main halo is similar to the mass of stars formed in satellites.  While the formation of stars in the central 1 kpc is reminiscent of observations of rapidly rotating pseudo bulges \citep{Kormendy1993}, we cannot draw any firm conclusions since the inner 1 kpc is unresolved in these simulations.  

Stars that form in the disk also contribute a large fraction of stars to the final spheroid, though a merger 6 Gyr into the simulation marks the end of the dominant contribution of disk stars to the spheroid.  We leave a detailed examination of how stars migrate from the disk to the spheroid for future work, but speculate that most of the migration is due to the tidal disruption caused by merging satellites given the reduction in disk contribution to the spheroid once the merging activity is complete.  We note that \citet{Debattista2004} showed that secular evolution accounted for some migration of stars from the disk to the spheroid in high resolution simulations of isolated disks.

Satellites are expected to contribute a large mass of stars to the spheroid since stars are tidally stripped into spheroidal orbits.  In the case of g15784's quiet merger history, the  mass of the spheroid formed in satellites is similar to that formed in the inner 1 kpc and the disk.  Finally, we note that the star formation in satellites is also largely confined to the unresolved central kpc similar to the main galaxy.

The large number of stars that form in the central kpc and disk echo the findings of \citet{Zolotov2009} and indicate that the predominant spheroid component is due in large part to stars that form in the main galaxy.  However, satellites contribute a significant mass of stars to the spheroid and disrupt a large fraction of stars out of the disk, so these simulations do not exclude traditional spheroid creation through satellite mergers and tidal stripping.  

\subsection{Density Profiles}
\label{sec:denprof}
\begin{figure}
\resizebox{9cm}{!}{\includegraphics{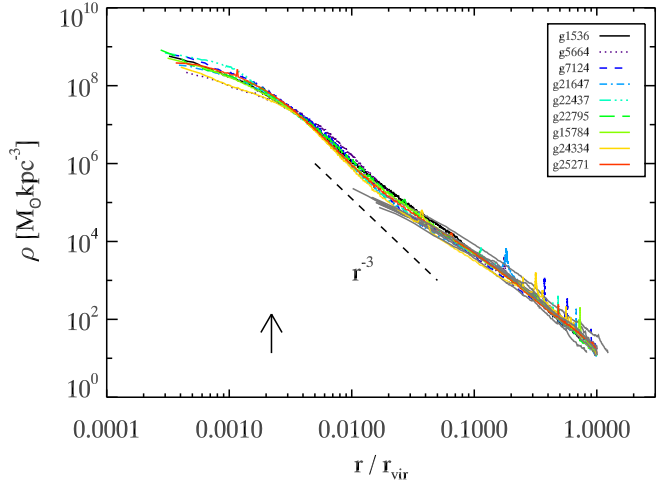}}
 \caption[Density Profiles]{Particles are sorted by radius and placed into 1000 bins each containing the same number of particles.  The radius of the bin is plotted as the mass weighted mean of the particle radii for each bin.  The total density profile (\textit{various colours}) for each halo is compared with the profile of the dark matter from the same simulations (\textit{thin grey}).  The arrow is placed at a radius of $2\epsilon_{soft}$, the radius inside which \citet{Power2003} determines density profiles can no longer be trusted.}
\label{fig:combden} 
\end{figure}

From the simulations, one can find the three dimensional distribution of all the matter in the galaxy as opposed to just that which is visible.  Figure \ref{fig:combden} shows the resulting radial density profile.  Particles are sorted by radius and placed into 1000 bins, each containing the same number of particles.  The dark matter only profiles for every galaxy are shown as the gray solid lines.  The dark matter and total density profiles are similar for each simulated galaxy.  The dark matter profiles follow a $r^{-3}$ power law or slightly shallower from the virial radius into 0.1$r_{vir}$ before growing shallower, similar to dark matter only runs \citep{NFW,Moore1998,Reed2005}.  The total profile continues a steady rise along the power law $r^{-2}$ into 1 kpc (about 0.005 $r_{vir}$) due to the stellar density before flattening out to a slope less than $r^{-1}$ inside 1 kpc.  Stars dominate the matter profile inside of 2 kpc.  

\subsection{Rotation Curves}
\begin{figure}
\resizebox{9cm}{!}{\includegraphics{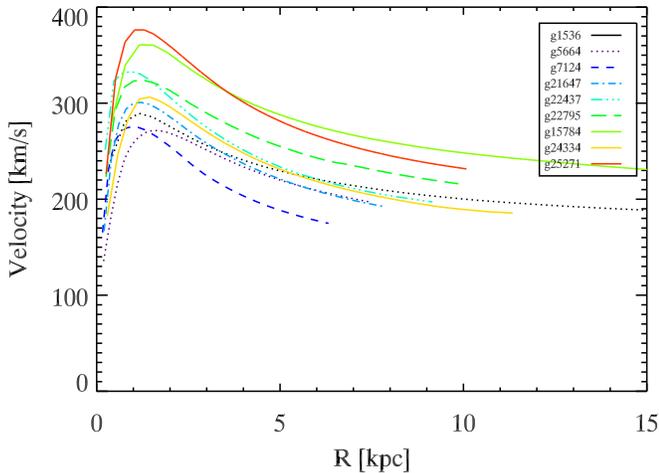}}
 \caption[Rotation Curves]{Rotation speed as a function of radius.  The circular velocity ($solid$) defined as $v_c=\sqrt{\frac{GM}{r}}$ in radial bins for each of our galaxies.  }
\label{fig:combrc} 
\end{figure}

Rotation curves provide a comparison between the density profile found in these simulations and observations.  Figure \ref{fig:combrc} shows the circular velocity, $v_c$ as a function of radius, where $v_c=\sqrt{\frac{GM}{R}}$.  Mock observations are not used because there is too much scatter in the star particle velocities at a given projected distance from the center of the galaxy.  

The excessive bulge illustrated in \S \ref{sec:bd} is again apparent in the rotation curves.  Rather than exhibiting flat rotation curves as are observed, the simulated rotation curves all have a peak at the center due to a large central concentration of mass.  \citet{Governato2007} showed that increased resolution without any changes to star formation can also limit the central concentration of matter as the lessened impact of two body interactions with halo dark matter particles minimizes angular momentum losses.  \citet{Governato2010} show how higher resolution and consequent modifications to the star formation threshold density can limit the mass concentration in simulations of slightly less massive galaxies.  

\subsection{Tully-Fisher}
\begin{figure}
\resizebox{9cm}{!}{\includegraphics{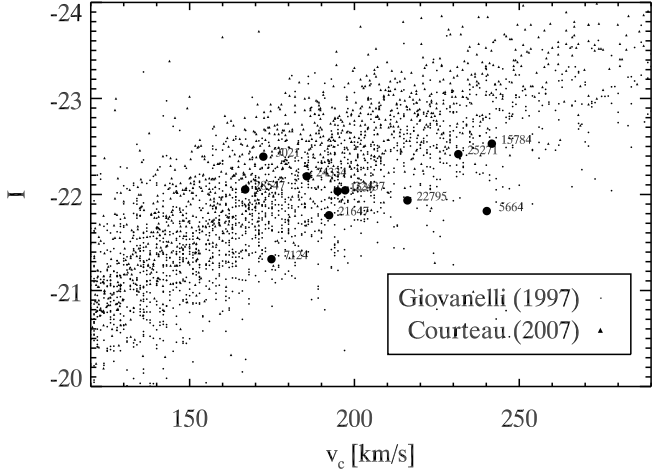}}
 \caption[Tully-Fisher Relationship]{I band absolute magnitude as a function of circular velocity at 3.5 disk scale lengths ($h$).  The large dots represent the simulated galaxies while the small dots are from the \citet{Giovanelli1997} sample of cluster galaxies and the triangles are from \citet{Courteau2007}.}
\label{fig:tf} 
\end{figure}

The Tully-Fisher (TF) relationship shown in Figure \ref{fig:tf} compares the luminosity of a galaxy with its rotation velocity.  Velocity indicates the dynamical mass for the inner parts of a galaxy, so the TF relationship indicates how bright a galaxy of a given mass should be.  Most galaxies in the range of masses that we simulated have flat rotation curves, so the radius at which the rotation velocity is measured is irrelevant.  In the simulations, however, the rotation curve rises sharply initially and the declines gradually with radius.  We follow the example of \citet{Governato2007} and measure the rotation velocity at 3.5$h$.  \citet{Governato2007} find in a resolution study that this is the radius at which velocity converges.  It is unclear from Figure \ref{fig:combrc} that the rotation curves are flat at this radius, but they are flatter than they are at smaller radii.  The I band magnitude for the galaxies generally matches the observations of the corresponding velocities, though the velocities remain slightly high.  \citet{Governato2010} showed that higher resolution and higher star formation density threshold produce galaxies with more slowly rising rotation curves, so higher resolution simulations of these galaxies is an obvious next step.

\subsection{Mass-to-Light Ratio}
\begin{figure}
\resizebox{9cm}{!}{\includegraphics{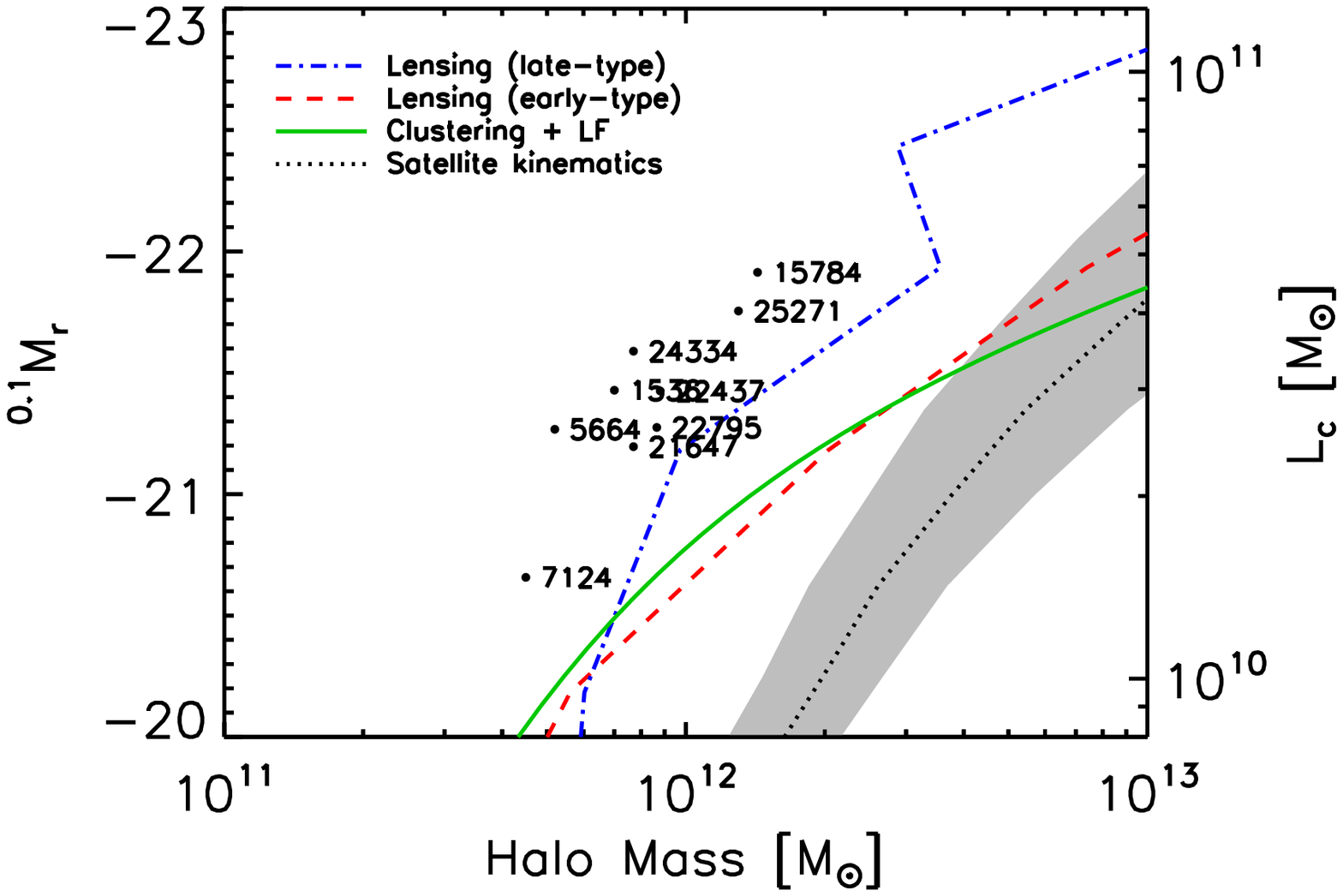}}
 \caption[M/L]{ Galaxy luminosity in the ${}^{0.1}r$ band plotted as a function of the
total halo mass including dark matter for the 9 simulated MUGS galaxies.
Observational measurements of the mass-luminosity relationship are shown
based on weak gravitational lensing \citep{Mandelbaum2006},
halo model fits to the galaxy clustering and luminosity function \citep{Cacciato2009}, and the kinematics of satellite galaxies \citep[][the measured dispersion is shown as the shaded region]{More2009}.}
\label{fig:ml} 
\end{figure}

The relationship between the luminosity of a galaxy and the mass of its
halo is a key observational question that provides constraints on
semi-analytic models and strongly affects the interpretation of observational samples with respect to theoretical predictions about halos.  In Figure \ref{fig:ml},
we have plotted the luminosity of our simulated galaxies versus their
total (dark plus baryonic) mass. The luminosities are K-corrected to the
${}^{0.1}r$ band (the $r$ band redshifted to $z=0.1$) using \textsc{kcorrect}
v4.1.4 \citep{Blanton2003} using the \textsc{sunrise}-generated SDSS $gri$
magnitudes in order to facilitate making comparison to the observations.

By construction, our galaxies inhabit a relatively narrow band in halo
mass, from $\sim 5 \times 10^{11}~M_{\sun}$ to $\sim 2 \times
10^{12}~M_{\sun}$.
Their luminosities cover a similar relative range in luminosity, from
$\sim 1.5 \times 10^{10}~L_{\sun}$ to $\sim 5 \times 10^{10}~L_{\sun}$,
for a typical total mass-to-light ratio of $\sim 30$ within the virial radius.

We have overplotted a variety of observational estimates of the
relationship
between central galaxy luminosity and the total halo mass onto Figure \ref{fig:ml}.
These observational techniques are: weak gravitational lensing
\citep[][blue dot-dashed line for exponential-dominated
late-type galaxies, and red dashed line for de~Vaucouleurs-dominated
early-type galaxies]{Mandelbaum2006}, a halo model simultaneously fit to galaxy clustering
and the luminosity function \citet[][solid green line]{Cacciato2009},
and the kinematics of satellite galaxies \citet[][dotted
black line for the mean and shaded gray region for the measured
spread in halo mass at a given luminosity]{More2009}. Note that while all of
these studies are based on SDSS data, the techniques use completely
independent
properties of the galaxies to determine the halo mass. We have adopted
$h=0.73$, as used in the simulations, to convert the $h$-independent
units used in these studies into physical units for direct comparison.

When comparing the observations to the simulations, it is important
to realize that in the simulations, the systems are selected based
on the halo mass, while the luminosity is an output of the simulations;
we have therefore plotted the halo mass as the independent variable
and the luminosity as the dependent variable in Figure \ref{fig:ml}. However,
in the observational studies, the galaxies are selected based on their
luminosity and the halo mass is what is measured; therefore, the
luminosity is the independent variable and the halo mass is the dependent
variable. This should be kept in mind when considering Figure \ref{fig:ml}, and
is why the weak lensing result around late-type galaxies appears
multi-valued.

The halo mass-luminosity relationship for the simulated galaxies has the
same shape and similar normalization as the observed relationship,
particularly when comparing to the \citet{Mandelbaum2006}
and \citet{Cacciato2009} results. The simulated galaxies are, however,
all more luminous than the observed galaxies at a given halo mass,
or equivalently the observed galaxies lie in more massive halos at a given
luminosity. This difference is no larger than the difference
seen between observational techniques, so it may not be significant. One
aspect of the simulated and observed samples that may be important
is that the simulated galaxies are constrained not to lie near large group
or cluster halos; given the mass dependence of halo bias, those halos
are themselves likely to be more massive. However, it seems unlikely that
this is the explanation given that the most isolated of the observational
samples is that of \citet{More2009}, with whom our results are the
most discrepant. A more likely explanation is that this is another symptom
of the overcooling problem, which turns too much of the gas in the system
into stars and therefore results in an overluminous galaxy.

\subsection{Star Formation Histories}
\begin{figure}
\resizebox{9cm}{!}{\includegraphics{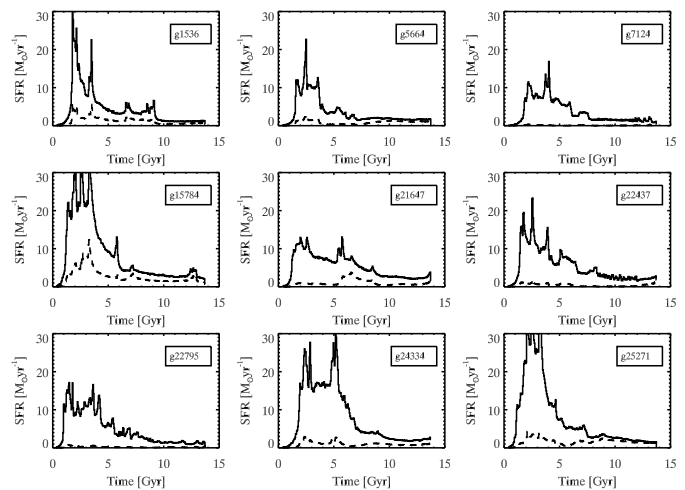}}
 \caption[Star Formation Histories]{The star formation rate plotted as a function of time for all the stars inside $r_{vir}$ ($solid$) and just stars that end up in the kinematically-defined disk ($dashed$).  Stars are sorted by their formation time into 50 Myr bins and the total mass of stars in that bin is divided by 50 Myr. }
\label{fig:combsfh} 
\end{figure}

One of the major ways to track the evolution of a galaxy is by examining when its stars formed.  Figure \ref{fig:combsfh} shows the star formation histories (hereafter, SFH) for each of our simulated galaxies.  Peaks in the SFH correspond to merger events which drive starbursts.  Mergers tidally disrupt disk gas and excite instabilities that cause gas overdensities \citep{Mihos1996, Springel2000, Cox2006}.  Following the merger peaks, the shape of the star formation history is an exponential increase followed by an exponential decline as the reservoir of gas is exhausted.  The galaxies in Figure \ref{fig:combsfh} that have the most and latest mergers correspond to those with the least well defined disks.

\subsection{Metallicity Evolution}
\begin{figure}
\resizebox{9cm}{!}{\includegraphics{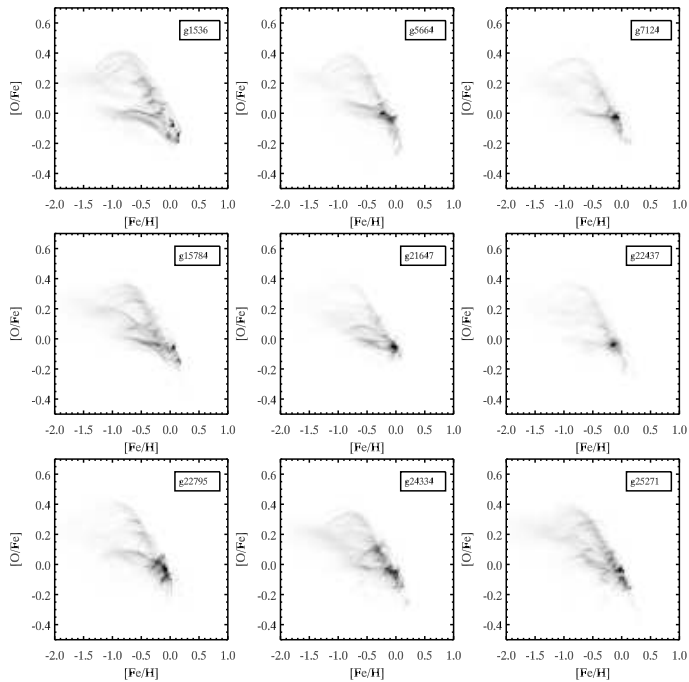}}
 \caption[Metallicities]{Metallicity distribution, [O/Fe] as a function of [Fe/H], of all the stars inside $r_{vir}$ of the galaxy. Stars are sorted into $100\times100$ bins, 0.03 wide in [Fe/H] and 0.01 high in [O/Fe].  Evolution progresses from low [Fe/H] and high [O/Fe] to solar (0.0) [Fe/H] and [O/Fe].  The multiple evolutionary tracks through this diagram correspond to different stellar structures that each have their own star formation history.}
\label{fig:combmet} 
\end{figure}

Trends in metallicity help track the star formation history of galaxies.  Specifically, Figure \ref{fig:combmet} shows the distribution of stars in [O/Fe] as a function of [Fe/H].  Evolution of metallicity proceeds from low [Fe/H] as stars form and produce iron that is ejected during supernova explosions.  Type II supernovae (SNII) occur on a shorter timescale than Type Ia, and they eject alpha elements such as carbon, oxygen, and silicon as well as iron.  We use oxygen to track the abundance of alpha elements in our simulations.  After stellar populations age for 40 Myr, Type Ia supernovae (SNIa) start to explode, ejecting few alpha elements, but large quantities of iron.  The range of [Fe/H] over which the [O/Fe] ratio remains high and constant indicates how many SNII explode before SNIa start contributing significant quantities of iron.  Thus, regions of active star formation like the Milky disk will generate long, high [O/Fe] plateaus before [O/Fe] decreases due to the influx of iron from SNIa \citep{Bensby2005,Reddy2006}.  Places where star formation is not as efficient like dwarf galaxies exhibit lower [O/Fe] at low [Fe/H] distinct from abundances of stars measured in the Milky Way's disk and halo \citep{Venn2004}.  Models of galaxy formation that couple N-body simulations with analytic prescriptions for the stellar content of satellite galaxies are able to reproduce this dichotomy \citep{Font06}.

Figure \ref{fig:combmet} shows that the majority of the stars in the simulations form with
solar metallicity and abundances. At low [Fe/H], stars form with both high
and low [O/Fe].  The peak [O/Fe] values are 0.4, which are lower than the observed abundances of the disk, which have been observed up to 0.6 \cite{Bensby2005,Reddy2006}.  This indicates a shortcoming of the simple power law fit used for oxygen enrichment.  The fit only allows type II supernovae from stars up to 40 M$_\odot$ to produce oxygen, so the chemical model does not capture chemical enrichment from more massive stars that produce more Oxygen compared to Iron \citep{Woosley95}.

Much like observational galactic archaeology, the distinct metallicity evolution tracks apparent in Figure \ref{fig:combmet} provide a useful alterative for discovering substructure inside galaxies as substructures exhibit stellar populations with different metallicity signatures.  We leave further discussion of such methods for future work.

\section{Conclusions}
We presented 9 galaxies from the McMaster Unbiased Galaxy Simulations simulated using N-body gravity and SPH.  The galaxies are selected from a mass range around the mass of the Milky Way and from isolated environments, but their selection was otherwise unbiased for factors such as accretion history and angular momentum.  

The galaxies were examined using the radiative transfer program \textsc{sunrise} to enable comparisons between the simulated galaxies and real galaxies in the observed plane.  The simulated galaxies have colors and magnitudes that compare well with a sample of inclination corrected isolated galaxies from SDSS, and in particular separate into the well-known red sequence and blue cloud.  However, both simulated populations tend too much towards the ``green valley'', indicating that they contain more old stars than blue cloud galaxies and more young stars than galaxies along the red sequence.

The surface brightness profiles of the simulated galaxies can all be fit with exponential disks combined with a central de Vaucouleurs $r^{1/4}$ law similar to real galaxies.  However, the proportion of bulge to total light (B/T) is higher than what is typically observed.  The B/T ratios are also high when the stars are decomposed into the spheroid and disk based on their kinematics.  There are no galaxies with a B/T fraction less than 0.5 whereas observed samples find many galaxies with B/T $<$ 0.5.  This result is similar to that found in many previous simulations.  We note that many of the recently formed stars that are classified as part of the spheroid form with orbits in the disk plane in the central regions of the disk.  We also note that most of the stars that comprise the spheroid formed \emph{in situ}, but we leave the question of how the stars may migrate from the disk to a spherical distribution for future work.

As to the question of why galaxies form with more or less spheroid, there seems to be a modest trend with accretion history.  We find that the largest disks (g1536 and g15784) form with a quiet merger history in which they had their last major merger prior to $z$=3, while the largest bulges all resulted from recent last major mergers.  There also appears to be a dependence on halo spin as g25271 has a relatively quiet merger history, but low $\lambda'$ and results in a significant bulge and red color.  Since all the galaxies used in MUGS fall in a limited mass range, these conclusions do not include the impact of mass.

We compare the brightness of the final galaxies with their mass at two different radii in the final output.  First, we compare our galaxies with the observed Tully-Fisher relationship that probes the amount of light a galaxy produces with the mass contained in its inner regions.  Since the final mass concentration of all the galaxies is too high, we use a rotation velocity from 3.5 disk scale lengths away from the galaxy center.  Previous studies have shown that this is the radius at which rotation velocities converge.  The galaxies are still slightly \emph{fainter} than the observed sample based on these inner velocities.  Second, we compare the brightness of our galaxies with observations of the whole halo mass derived using a number of different methods.  In each case, the simulated galaxies are \emph{brighter} than  comparable observed galaxies at the same mass.  This indicates that too many stars form in the simulation.  The high central velocities used in the Tully-Fisher relationship indicate that mass gets too concentrated at the centers of the halos and while the galaxies are fainter than the observed TF relationship, the lack of resolution makes it difficult to determine whether the amount of stars formed is too many or too few.

We are thus left with the challenge of creating more realistic simulations in order to obtain more accurate insights into how the important physical processes involved in galaxy formation result in the observed population of galaxies.  Fortunately, there has been much recent work that guides the way forward.  It has been shown repeatedly that higher resolution makes better disks \citep{Governato2004,Governato2007}.  More recently, it has been shown that high resolution combined with clustered star formation can remove central density concentrations from dwarf galaxies \citep{Mashchenko2008,Governato2010}. 

While these simulations open many possibilities for expanding our understanding of how galaxies form, they also show that there is much work left to be done before we can claim to have simulated a sample of galaxies that compares well with real ones.

\section*{Acknowledgments}

This paper makes use of simulations performed as part of the SHARCNET Dedicated
Resource project: “MUGS: The McMaster Unbiased Galaxy Simulations Project” (DR316,
DR401, and DR437).  We would like to thank Allison Sills, Bill Harris, and Victor Debatista for useful conversations.  We would also like to thank Rok Ro$\check{s}$kar and Peter Yoachim for helpful IDL code that contributed to this project.  As should be apparent from the bibliography, much of this work was inspired by Fabio Governato.  GS is a Fellow of the Jeremiah Horrocks Institute and did the bulk of this work as a CITA National Fellow.  JB and HC were supported by NSERC grant PHY-0205413.  JW was supported NSERC grant.

\bibliographystyle{mn2e}
\bibliography{references}

\clearpage

\end{document}